# Revealing Invisible Scattering Poles via Complex Frequency Excitations

Deepanshu Trivedi[1], Arjuna Madanayake[1], and Alex Krasnok[1,2*]

[1]Department of Electrical and Computer Engineering, Florida International University, Miami, Florida 33174, USA

[2]Knight Foundation School of Computing and Information Sciences, Florida International University, Miami, FL 33199, USA

*To whom correspondence should be addressed: akrasnok@fiu.edu

## Abstract

*Recent research in light scattering has prompted a re-evaluation of complex quantities, particularly in the context of complex frequency signals, which exhibit exponential growth or decay unlike traditional harmonic signals. We introduce a novel approach using complex frequency signals to reveal hidden or invisible poles—those with predominantly imaginary components—previously undetected in conventional scattering experiments. By employing a carefully tuned complex frequency excitation method, we demonstrate the efficient conversion of non-oscillating fields into oscillating ones. This effect is shown in both RF and optical domains, specifically within the C-band infrared spectral range, which is crucial for communications. This study enhances the theoretical framework of wave interactions in photonic systems, paving the way for innovative applications in invisibility cloaking, advanced photonic devices, and the future of optical communication and quantum computing.*

Negative numbers first appeared in ancient China (~200 BCE) and India (7th century CE) but faced skepticism in Europe until the 17th century due to their perceived lack of real-world meaning. Arnauld's paradox, for instance, argued that if negative numbers existed, then (−1)/1 should equal 1/(−1), suggesting a nonsensical equivalence between smaller-to-larger and larger-to-smaller ratios. This confusion persisted until scientific advancements, particularly in the 18th century, began to change perspectives. Founding Father Benjamin Franklin played a pivotal role in this shift. In the 1740s, Franklin proposed the concept of positive and negative electric charges, providing the first concrete example of a negative quantity in nature—the negative electric charge of the electron. This was a groundbreaking idea, paving the way for Maxwell's theory of electromagnetism in the 19th century, which fundamentally relied on the existence of negative charges [1].

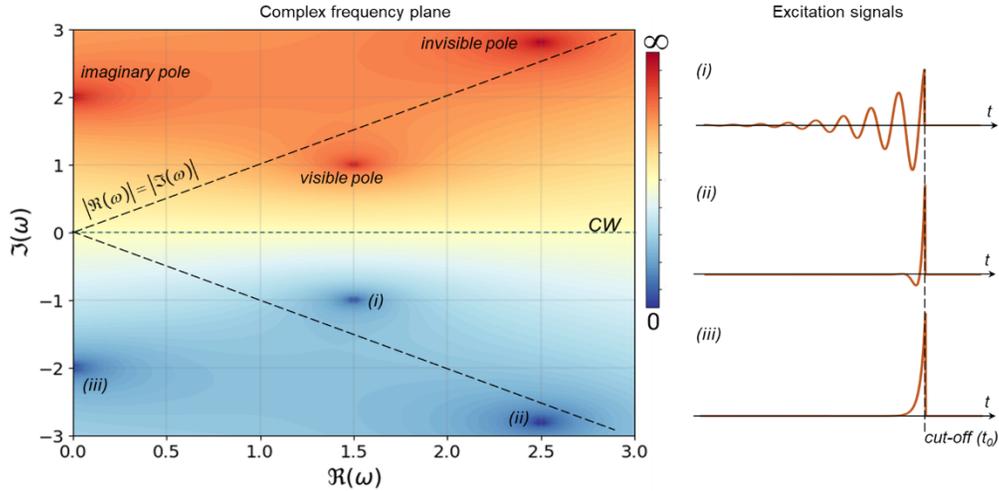

*Figure 1.* Left: Response function of an arbitrary system in the complex frequency plane. "Visible poles" radiate energy as oscillating fields. "Invisible poles" rapidly decay energy. "Imaginary poles" decay without oscillation. Right: the corresponding excitation signals.

Today, a similar skepticism surrounds complex numbers, often perceived as abstract mathematical constructs. However, they are fundamental to quantum mechanics. The Schrödinger equation, which governs quantum mechanics, inherently involves the imaginary unit. This inclusion is not arbitrary; without the imaginary unit, the equation would become second-order in time, necessitating two initial conditions, which does not align with the physical reality of quantum systems. The wave function, which describes the state of a quantum system, is inherently complex, and this complexity is essential for accurately representing phenomena like superposition and interference. Techniques such as quantum tomography even allow for the indirect measurement of these complex wave functions, underscoring their real-world significance [2–4]. Thus, complex numbers are indispensable for accurately modeling and predicting both quantum and photonic behavior.

Complex frequency (CF) signals, characterized by their ability to exhibit exponential growth or decay, represent a significant departure from traditional harmonic signals. These signals have become increasingly important in understanding and manipulating systems with losses or gains, where natural frequencies become complex and reveal behaviors such as exponential decay or growth [5]. This reevaluation of CF signals has profound implications for advanced light scattering methods, which are crucial for controlling wave propagation in photonics and quantum systems [6–8]. By leveraging CF signals, researchers can achieve unprecedented control over wave behaviors, enabling innovations in areas such as invisibility cloaking and the development of novel photonic devices. The ability to manipulate light at the quantum level using CF signals opens new avenues for both scientific exploration and technological advancement [9–12]. These advancements promise to revolutionize our understanding of wave interactions, leading to breakthroughs in invisibility technologies, the creation of new classes of photonic devices, and enhanced capabilities in controlling light within quantum systems.

Note that just like a harmonic signal, a complex frequency signal is a mathematical abstraction, as it must start and stop at specific moments. However, similar to harmonic functions, complex frequency functions are valuable and can be generated in a laboratory using various methods across RF [13], microwave [14], acoustics and elastodynamics [15], and optics [16].

Figure 1 shows a response function of an arbitrary system in the complex frequency plane. This extension of the response function to the complex frequency plane helps verify system causality and relates to the Laplace transformation [17]. If the system were closed and lossless, all poles (system eigenmodes) and zeros (no response) would lie on the real frequency axis. This is where continuous wave (CW) frequencies are located, representing the system's discrete bound states. However, because the system is open, these poles shift to the complex plane. We refer to poles with an imaginary frequency part smaller than the real part as "visible poles." These poles can radiate energy as an oscillating field when excited. To excite these poles one use the CF excitations (Figure1(i)). On the other hand, as we demonstrate below in this work, there are "invisible poles," whose imaginary frequency part is greater than the real part. This means the energy from these poles decays so quickly that the field cannot complete even one oscillation. In the theory of oscillations, these are known as overdamped modes [18]. Typically, these modes result from visible poles being pushed into the complex plane over the $|\Re(\omega)|=|\Im(\omega)|$ line by dissipative losses, like friction in a pendulum.

In this work, we show that invisible poles can exist even in Hermitian systems, which follow time-reversal symmetry. This raises the question: Can these invisible poles be excited, and do they affect scattering in such systems? As we show below, these poles play a crucial role in scattering. They can be excited by signals that grow and barely make a single oscillation (Figure 1(ii)). Additionally, there are another type of eigenmodes located on the imaginary frequency axis, which we call "imaginary poles." These poles are of a completely imaginary nature and can be excited by a type of signal known as an *imaginary* frequency signal, which only grows without oscillation (Figure 1(iii)). We discuss the feasibility and importance of such signals in what follows. Remarkably, we also demonstrate that a non-oscillating CF pulse can be effectively converted into oscillating signals. In this work we use the $e^{j\omega t}$ time convention and $j$ is the imaginary unit, $j = \sqrt{-1}$.

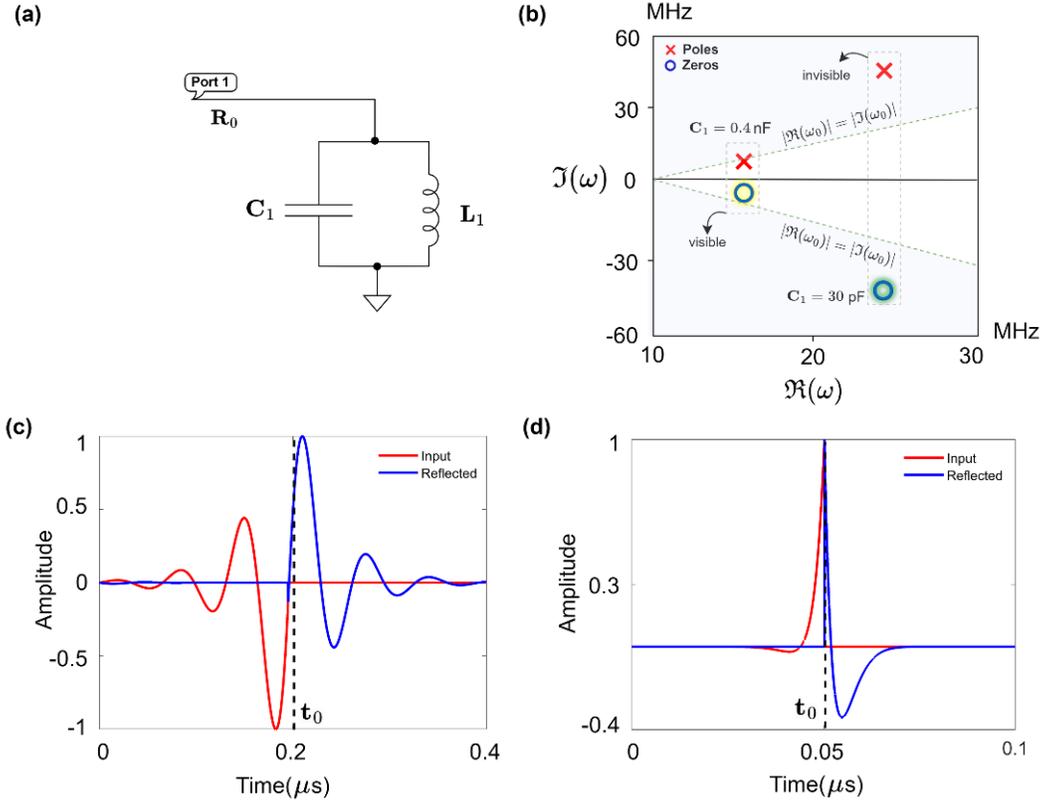

**Figure 2.** *Lossless LC resonant circuit with parameters $L_1$ = 0.25 μH and $C_1$ = 0.4 nF (visible pole) or $C_1$ = 30 pF (invisible pole). (b) Configuration showing the visible pole at $\tau\omega_0 = 0.5$ and the invisible pole at $\tau\omega_0 = 1.83$. (c) Simulation results demonstrating the excitation of the visible pole. (d) Transition to the "invisible regime" at $\tau\omega_0 = 1.83$, showing significant changes in scattering and energy radiation without oscillation.*

When designing photonic systems, we often use the lumped element circuit approach for its simplicity and effective analysis techniques. Using this method, we have explored light scattering anomalies in RLC circuits and observed interesting effects [19]. Let us begin with examination of a lossless LC resonant circuit illustrated in Figure 2(a). The circuit exhibits a single mode with a real angular eigenfrequency, $\omega_0 = 1/\sqrt{LC}$. The circuit is connected to a transmission line with $R_0 = 50$ Ω characteristic impedance. The reflection coefficient is $r(\omega) = \dfrac{Z_L(\omega) - R_0}{Z_L(\omega) + R_0}$, and $Z_L(\omega) = R_L(\omega) + jX_L(\omega)$ being the load input impedance, $X_L$ is an effective frequency-dependent reactance and $R_L$ is the load resistance [20]. In the purely reactive regime ($R_L = 0$), the load reactance is $X_L(\omega) = \omega L \dfrac{\omega_0^2}{\omega_0^2 - \omega^2}$ [13]. Hence, the complex reflection coefficient is

$$r(\omega) = -\dfrac{(\omega^2 - \omega_0^2) + j(\tau\omega_0^2)\omega}{(\omega^2 - \omega_0^2) - j(\tau\omega_0^2)\omega},$$

where $\tau = L / R_0$. Solving this equation for $r(\omega) = \infty$ yields frequencies of poles $\omega_p = j\dfrac{\omega_0}{2}\left(\tau\omega_0 \mp \sqrt{\tau^2\omega_0^2 - 4}\right)$. Similarly, for the zeros we get,

$$\omega_z = -j\dfrac{\omega_0}{2}\left(\tau\omega_0 \pm \sqrt{\tau^2\omega_0^2 - 4}\right).$$

The response of this circuit is defined by the value of the value of the coupling parameter $\tau\omega_0$. The case $\tau\omega_0 < 2$ is of utmost importance for applications, because, in this case, both the zero and the pole have nonzero real frequencies, making them accessible to waves and signals. In the extreme case where there is no coupling (when the coupling parameter is zero, $\tau\omega_0 = 0$), the convergence of the pole and zero on the real axis occurs, leading to the formation of a discrete bound state characterized by a diverging quality factor. As the coupling of the circuit to the port increases, the pole and zero move further apart in the complex plane.

Let's consider the configuration where $\tau\omega_0 = 0.5$, which corresponds to the "visible pole" in Figure 2(b). The circuit parameters are $L_1 = 0.25$ µH and $C_1 = 0.4$ nF. This pole can be excited through the corresponding zero by an exponentially growing signal, as shown in Figure 2(c). For the simulation, we use PathWave ADS with its transient/convolution solver for SPICE-type transient time-domain analysis. The results show that when we excite the pole by matching the excitation to its zero (red curve), there is no reflected signal (blue curve), and all the energy sent to the system gets perfectly trapped in the resonator. When the excitation stops, this visible pole radiates at its complex frequency. Due to the Hermitian nature of the system, the response is time-reversed.

Now, let us examine what happens when we increase the parameter $\tau\omega_0$ to 1.83 by changing $C_1$ to 30 pF. In this case, the pole and the zero cross the $|\Re(\omega)|=|\Im(\omega)|$ border line, entering the invisible regime where the real frequency is less than the imaginary frequency. This transition significantly affects the scattering, as shown in Figure 2(d). When we excite the pole via its zero using the red input signal, there is no scattering before the cut-off time $t_0$. After the excitation stops, all the energy radiates back without oscillation. This illustrates the transition from the visible to the invisible pole scattering regime. When excited at the real frequency axis, this pole does not manifest itself at all. It is important to note that the system is Hermitian, meaning this invisible pole is not due to the conventional overdamped mechanism.

Thus, the invisible pole can be excited and revealed in scattering as a decaying signal without oscillations. Even though its response has a real frequency, it cannot complete a full oscillation before it decays. However, this pole can accumulate energy when excited through the associated zero, raising an intriguing question: Can this pole interact with the usual visible pole, and how can this interaction be observed?

Figure 3(a) shows a circuit with two resonators. The leftmost resonator is coupled directly to the port and has an invisible pole. The other resonator has a visible pole. We tuned the coupling strength $C_c$ so that the poles have close real frequencies, as shown in Figure 3(b). We tested this circuit by exciting it via the zero corresponding to the visible pole, Figure 3(c). The result shows no scattering before cut-off and a reflected signal after cut-off. Remarkably, the output signal is not a time-reversed version of the input signal, especially right after the cut-off. This is due to interference from both poles, indicating that the presence of the invisible pole affects the total response of the system. The inset shows the fast Fourier transform (FFT) spectrum of the incident signal before the cut-off and the reflected signal after the cut-off. Both signals are centered around 2 MHz.

An even more interesting and somewhat contradictory result is observed when we excite the circuit with a non-oscillating field at the CF corresponding to the zero of the invisible pole, as shown in Figure 3(d). In this case, after the cut-off, the circuit produces an oscillating reflected signal. The FFT spectrum of this signal reveals the real frequency of the visible pole, 2 MHz. This demonstrates that the energy from the invisible pole can be coupled to the visible pole and then radiated as an oscillating field via the visible pole.

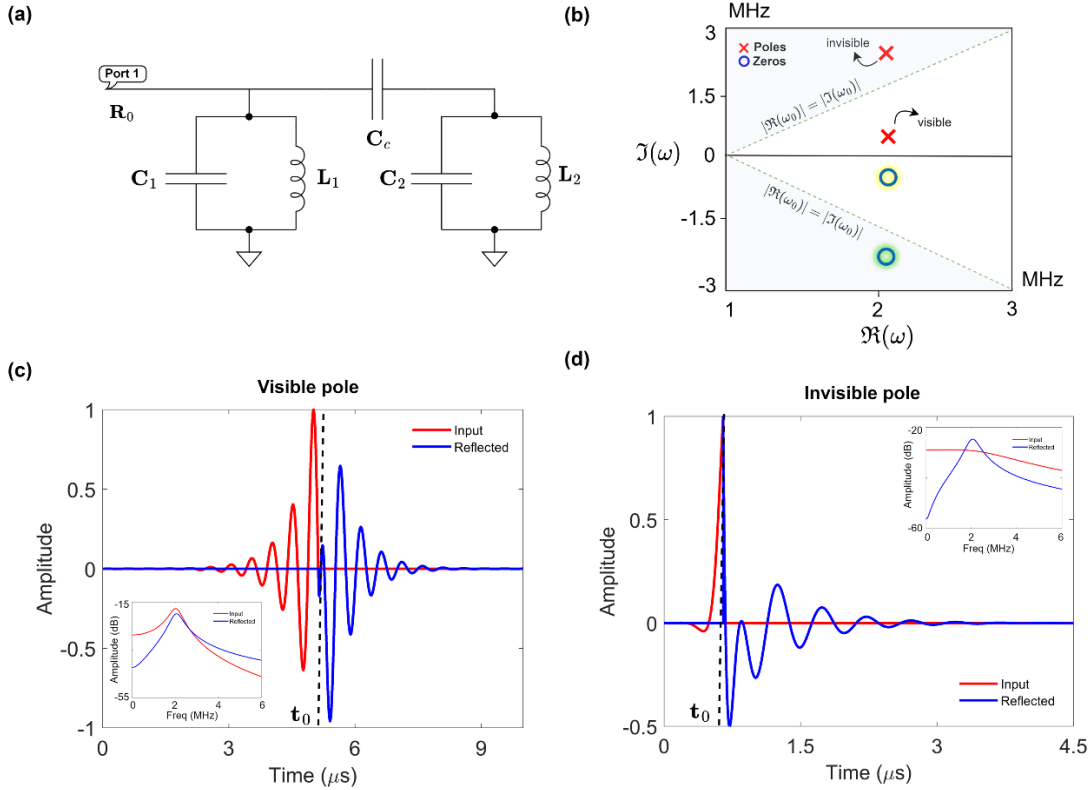

**Figure 3.** *(a) Circuit with two resonators: the left (invisible pole) resonator is coupled to the port, and the right has a visible pole. (b) Coupling strength $C_c$ is tuned to match real frequencies of both poles. (c) Excitation via the visible pole zero shows no scattering before cut-off and a reflected signal after cut-off. The output signal, especially right after the cut-off, is not time-reversed due to pole interference. The inset shows the FFT spectrum centered around 2 MHz for both signals. (d) Excitation at the invisible pole's zero complex frequency results in an oscillating reflected signal after cut-off. The FFT spectrum reveals the visible pole's 2 MHz frequency, demonstrating energy coupling from the invisible to the visible pole. The efficiency of the non-oscillatory-to-oscillatory signal conversion is ~31%.*

To measure the efficiency of this conversion, we directly integrate the energy of the incident signal in the time domain and compare it to the energy of the oscillatory part of the reflected signal (excluding the initial spike caused by the fast-decaying radiation from the invisible pole). By taking the ratio of these two energies, we find that the efficiency of this conversion is 31%. This conversion efficiency depends significantly on the coupling of the poles in the complex plane. For example, in the Supplementary Materials (Figure S1), we provide similar calculations for a similar circuit with detuned poles, meaning the poles have different real frequencies. The efficiency analysis for that circuit shows an efficiency of about 20%, which is lower than the efficiency in the matched case.

This conversion of a non-oscillating field into an oscillating field at the frequency controlled by the circuit's parameters is a general effect that can be observed in various systems, not just radio frequency circuits. It has significant technological applications. For example, it allows energy delivery to cryogenic circuits operating at millikelvin temperatures. The traditional approach uses coaxial cables to connect these circuits to RF equipment, which also brings in room temperature noise, requiring additional measures to prevent noise from reaching the cryogenic circuits [21]. Our technology enables the excitation and control of these circuits using simple wires, eliminating these disadvantages.

Remarkably, this visible to invisible pole coupling manifest itself also with purely imaginary poles sitting at the imaginary frequency axis. In Supplementary Materials, we investigate a 2-port circuit, Figure S2(a). The analytical analysis of the reflection coefficient in the complex frequency plane for this circuit reveals that the poles are not only in the complex plane but also on the imaginary axis, Figure S2(b). In this 2-port regime, the poles and zeros of the reflection are not complex conjugates due to the radiation to the second port. However, the 2x2 scattering matrix of this circuit has poles and zeros that are complex conjugates. Figures S2(c) and S2(d) display the reflected (c) and transmitted (d) signals when the circuit is excited by a *purely imaginary pulse* (with the excitation zero highlighted in green in Figure S2(b)). The computed conversion efficiency from non-oscillating to oscillating radiation is approximately 38%. This is slightly higher than in the single-port regime because the circuit now has two ports, providing two radiation channels, although the poles are largely detuned.

To show how common these effects are, we look at another system in the optical realm. Figure 4(a) shows an optical integrated photonics structure. It has a single-mode waveguide with a grating coupler for excitation and readout. The waveguide is side-coupled to a single-mode disk resonator with mode amplitude $a_1$ and eigenfrequency $\omega_{01}$. The coupling rate between the waveguide and the resonator is $\gamma_e$. This first resonator is then coupled to a second disk resonator via a near-evanescent field. The second resonator has a mode amplitude $a_2$ and eigenfrequency $\omega_{02}$. This coupling is described by the coupling strength $g$.

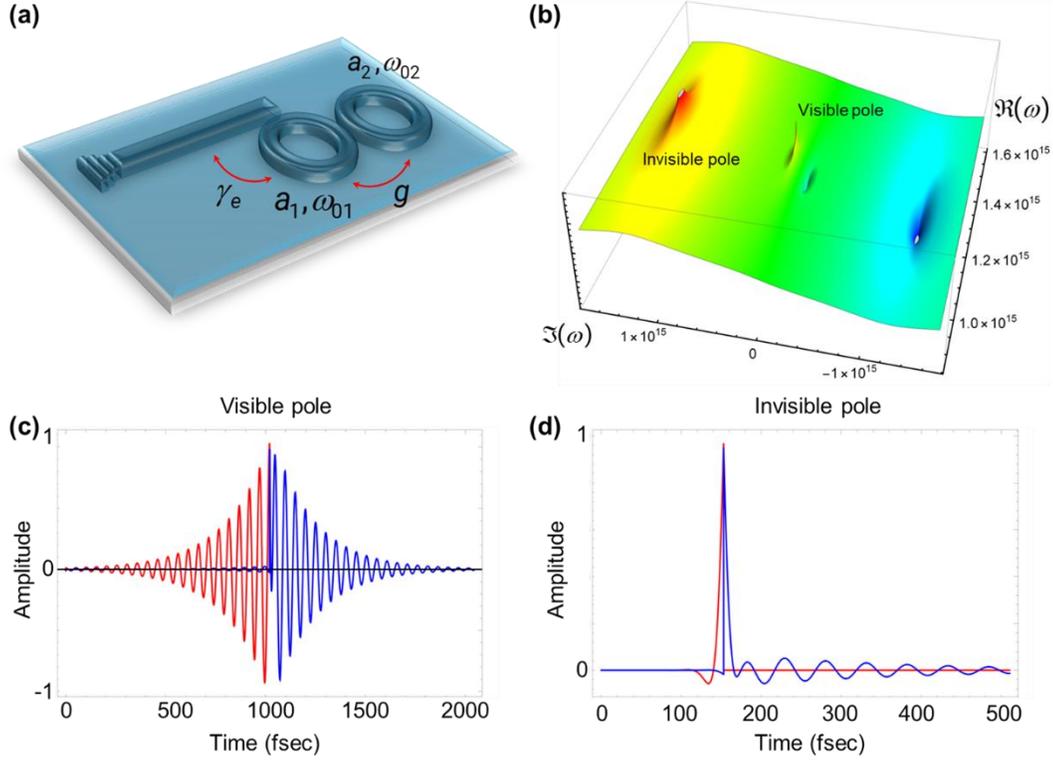

*Figure 4. (a) Schematic of the optical integrated photonics structure, featuring a single-mode waveguide with a grating coupler for excitation and readout, side-coupled to a single-mode disk resonator (mode amplitude a₁, eigenfrequency ω₀₁) and further coupled to a second disk resonator (mode amplitude a₂, eigenfrequency ω₀₂). (b) Scattering coefficient plotted in the complex frequency plane. (c) Reflected signal when the system is excited via the visible pole, with the reflected signal closely resembling the time-reversed excitation. (d) Reflected signal when the system is excited via the invisible pole, demonstrating the conversion to an oscillating field radiated through the visible pole.*

We focus on C-band infrared optics with a real frequency of 196 THz. To analyze the system, we use temporal coupled mode theory [22–25], which is detailed in the Supplementary Materials. There, we also derive the solution for the reflection coefficient in the steady-state regime,

$$R(\omega_r, \omega_i) = \left| -1 + 2\gamma_{e1} \det(M)^{-1} (i\omega_r - \omega_i - i\omega_{02} + \gamma_{o2}) \right|^2, \tag{1}$$

where $\det(M) = (i\omega_r - \omega_i - i\omega_{01} + \gamma_{o1} + \gamma_{e1})(i\omega_r - \omega_i - i\omega_{02} + \gamma_{o2}) - g^2$. Figure 4(b) shows the scattering coefficient plotted in the complex frequency plane. For this analysis, we chose the following parameters: intrinsic (unloaded) decay time $\tau_{o1,2} = 1/\gamma_{o1,2}$ for both the first and second resonators is 2000 T (corresponds to an intrinsic Q-factor of about 6300, a typical and feasible value), coupling time of the cavity to the waveguide $\tau_{e1} = 1/\gamma_{e1}$ is 0.15 T. Here, T is the period of one oscillation at the frequency of 196 THz. The mode coupling strength is $g = i\omega_0/5$. These parameters ensure that the system has both invisible and visible poles and that the poles are strongly coupled, sharing the same real frequency. Derivation of Eq.(1), of the determinant det(*M*) along with the approach description are given in the Supplementary Materials.

We use the differential equations from the temporal coupled mode theory in the time domain to calculate the mode amplitudes. Then, we compute the reflected signal over time when the system is excited by a corresponding complex frequency signal. Figure 4(c) shows the system being excited via the visible pole. In this case, the reflected signal is nearly a time-reversed version of the excitation signal. On the other hand, when the system is excited via its invisible pole by a non-oscillating field, it converts to an oscillating field that is radiated through the visible pole, as shown in Figure 4(d). The recent experimental demonstration of complex frequency effects in the optical infrared domain confirms the feasibility of testing these effects experimentally [16].

Finally, let's discuss two important aspects of the observed effects. First, these effects can be understood through signal analysis and filter theory. When a signal is quickly excited and then stopped, it is localized in time and can be Fourier transformed. This transformation, as shown in the insets of Figures 2 and 3, reveals that these signals are very broad in the frequency domain. The second resonator acts as a Purcell filter [26], which filters out specific frequencies and enhances coupling to the frequency of the "visible pole." While this explanation is valid, our analysis provides deeper insights into interactions in the complex frequency plane and shows how to improve the coupling between non-oscillating and oscillating waves.

Secondly, let's discuss the physical limitations of complex frequency signals. Signals with complex frequencies require a wide spectral bandwidth and high energy density because they show exponential growth or decay over time. Practical systems have limits on the power and energy that can be transmitted or converted into such signals. High energy density over short time intervals can cause overheating or damage to materials and components. Additionally, real-world environments cause attenuation and dispersion, limiting the frequency range and feasibility of using purely imaginary frequencies. High-frequency components also suffer significant losses due to environmental interactions, further restricting their practical use.

In conclusion, this work highlights the significance of complex frequency scattering effects in photonics, demonstrating their essential role in accurately describing and manipulating physical systems. It has been shown that certain systems possess invisible or imaginary poles, which are not detectable through conventional scattering experiments. By exciting the system with complex frequency signals, these hidden poles can be revealed. Furthermore, it has been demonstrated that these insights facilitate the efficient conversion of non-oscillating fields into oscillating fields. These findings enhance the understanding of wave interactions in photonic systems and pave the way for innovative applications, including invisibility and advanced photonic devices. The study provides deeper insights into interactions in the complex frequency plane. Additionally, the physical limitations of complex frequency signals have been discussed, highlighting the challenges posed by wide spectral bandwidth requirements, high energy density, attenuation, and dispersion in practical systems.

**Supplementary Materials**

The supplementary materials detail the analysis of conversion efficiency in resonant circuits, focusing on pole coupling in the complex frequency plane. Key findings include a 31% efficiency in a matched circuit, 20% in detuned poles, and 38% in a 2-port circuit. Temporal coupled mode theory and steady-state reflection coefficients are derived and analyzed.

**Acknowledgments**

The authors also thank the ECE department of Florida International University. This work is partially supported by DOE SBIR grant #DE-SC0024806.

**AUTHOR DECLARATIONS**

**Conflict of Interest**

The authors declare no competing financial interest.

**DATA AVAILABILITY**

The data that support the findings of this study are available from the corresponding authors upon reasonable request.

**REFERENCES**

[1]    J. C. Maxwell, *A Treatise on Electricity and Magnetism* (Oxford University, 1873).
[2]    D. Leibfried, D. M. Meekhof, B. E. King, C. Monroe, W. M. Itano, and D. J. Wineland,

*Experimental Determination of the Motional Quantum State of a Trapped Atom*, Phys. Rev. Lett. **77**, 4281 (1996).

[3] G. Breitenbach, S. Schiller, and J. Mlynek, *Measurement of the Quantum States of Squeezed Light*, Nature **387**, 471 (1997).

[4] T. Jullien, P. Roulleau, B. Roche, A. Cavanna, Y. Jin, and D. C. Glattli, *Quantum Tomography of an Electron*, Nature **514**, 603 (2014).

[5] A. Krasnok, D. Baranov, H. Li, M.-A. Miri, F. Monticone, and A. Alú, *Anomalies in Light Scattering*, Adv. Opt. Photonics **11**, 892 (2019).

[6] D. R. Bohren, Craig F, Huffman, *Absorption and Scattering of Light by Small Particles* (Wiley-VCH Verlag GmbH, Weinheim, Germany, Germany, 1998).

[7] J. D. Jackson, *Classical Electrodynamics* (John Wiley & Sons, 2021).

[8] M. O. Scully and M. S. Zubairy, *Quantum Optics* (Cambridge University Press, 1997).

[9] E. Rephaeli, J.-T. Shen, and S. Fan, *Full Inversion of a Two-Level Atom with a Single-Photon Pulse in One-Dimensional Geometries*, Phys. Rev. A **82**, 033804 (2010).

[10] S. Zhang, C. Liu, S. Zhou, C.-S. Chuu, M. M. T. Loy, and S. Du, *Coherent Control of Single-Photon Absorption and Reemission in a Two-Level Atomic Ensemble*, Phys. Rev. Lett. **109**, 263601 (2012).

[11] M. Bader, S. Heugel, A. L. Chekhov, M. Sondermann, and G. Leuchs, *Efficient Coupling to an Optical Resonator by Exploiting Time-Reversal Symmetry*, New J. Phys. **15**, 123008 (2013).

[12] J. Wenner et al., *Catching Time-Reversed Microwave Coherent State Photons with 99.4% Absorption Efficiency*, Phys. Rev. Lett. **112**, 210501 (2014).

[13] A. V. Marini, D. Ramaccia, A. Toscano, and F. Bilotti, *Perfect Matching of Reactive Loads Through Complex Frequencies: From Circuital Analysis to Experiments*, IEEE Trans. Antennas Propag. **70**, 9641 (2022).

[14] T. Delage, O. Pascal, J. Sokoloff, and V. Mazières, *Experimental Demonstration of Virtual Critical Coupling to a Single-Mode Microwave Cavity*, J. Appl. Phys. **132**, 153105 (2022).

[15] G. Trainiti, Y. Ra'di, M. Ruzzene, and A. Alù, *Coherent Virtual Absorption of Elastodynamic Waves*, Sci. Adv. **5**, eaaw3255 (2019).

[16] J. Hinney, S. Kim, G. J. K. Flatt, I. Datta, A. Alù, and M. Lipson, *Efficient Excitation and Control of Integrated Photonic Circuits with Virtual Critical Coupling*, Nat. Commun. **15**, 2741 (2024).

[17] Harald J W Muller-Kirsten, *Electrodunamics: An Inroduction Including Quantum Effects* (World Scientific, Singapore, 2004).

[18] A. Blais, A. L. Grimsmo, S. M. Girvin, and A. Wallraff, *Circuit Quantum Electrodynamics*, Rev.

Mod. Phys. **93**, 025005 (2021).

[19] D. Trivedi, A. Madanayake, and A. Krasnok, *Anomalies in Light Scattering: A Circuit Model Approach*, Phys. Rev. Appl. **in press**, (2024).

[20] D. M. Pozar, *Microwave Engineering, 4th Edition* (John Wiley & Sons, Inc., 2011).

[21] J. C. Bardin, D. H. Slichter, and D. J. Reilly, *Microwaves in Quantum Computing*, IEEE J. Microwaves **1**, 403 (2021).

[22] Y. Ra'di, A. Krasnok, and A. Alù, *Virtual Critical Coupling*, ACS Photonics **7**, 1468 (2020).

[23] W. Suh, Z. Wang, and S. Fan, *Temporal Coupled-Mode Theory and the Presence of Non-Orthogonal Modes in Lossless Multimode Cavities*, IEEE J. Quantum Electron. **40**, 1511 (2004).

[24] S. Fan, W. Suh, and J. D. Joannopoulos, *Temporal Coupled-Mode Theory for the Fano Resonance in Optical Resonators*, J. Opt. Soc. Am. A **20**, 569 (2003).

[25] H. A. Haus, *Waves and Fields in Optoelectronics*, Prentice-H (Prentice-H, Englewood Cliffs, NJ, Englewood Cliffs, NJ, 1984).

[26] C. Axline, M. Reagor, R. Heeres, P. Reinhold, C. Wang, K. Shain, W. Pfaff, Y. Chu, L. Frunzio, and R. J. Schoelkopf, *An Architecture for Integrating Planar and 3D CQED Devices*, Appl. Phys. Lett. **109**, 25 (2016).

# Supplementary Materials:

# Revealing Invisible Scattering Poles via Complex Frequency Excitations

Deepanshu Trivedi[1], Arjuna Madanayake[1], and Alex Krasnok[1,2*]

[1]Department of Electrical and Computer Engineering, Florida International University, Miami, Florida 33174, USA

[2]Knight Foundation School of Computing and Information Sciences, Florida International University, Miami, FL 33199, USA

*To whom correspondence should be addressed: akrasnok@fiu.edu

## 1. Analysis of the single resonator circuit in Figure 2(a)

Let us examine a lossless LC resonant circuit illustrated in **Figure 2(a)**. The circuit exhibits a single mode with a real angular eigenfrequency, $\omega_0 = 1/\sqrt{LC}$. When excited, the current and voltage oscillate at this real eigenfrequency without decay, with the electric field in the capacitor ($C$) and the magnetic field in the inductor ($L$) being phase-offset by $\pi/2$. When the circuit is connected to a transmission line, it may lose energy through reflection. The reflection coefficient is defined as $r(\omega) = \frac{Z_L(\omega) - R_0}{Z_L(\omega) + R_0}$, with $R_0$ being the characteristic impedance of the port transmission line ($R_0 = 50$ Ω) and $Z_L(\omega) = R_L(\omega) + jX_L(\omega)$ being the load (resonant circuit) input impedance, $j = \sqrt{-1}$, $X_L(\omega)$ is an effective frequency-dependent reactance, seen at the load terminals and $R_L$ is the load resistance(*1*).

In the purely reactive load regime ($R_L = 0$) and for the parallel connection of the inductor and capacitor, the load reactance ($X_L$) is $X_L(\omega) = \omega L \frac{\omega_0^2}{\omega_0^2 - \omega^2}$ (*2*). Thus, this circuit allows a fully analytical treatment with the following results for the complex reflection coefficient: $r(\omega) = -\frac{(\omega^2 - \omega_0^2) + j(\tau\omega_0^2)\omega}{(\omega^2 - \omega_0^2) - j(\tau\omega_0^2)\omega}$, where $\tau = L/R_0$. The case of the series connection of L and C differs only in the sign and value of $\tau$. Solving this equation for $r(\omega) = 0$ yields frequencies of zero reflection, $\omega_z = -j\frac{\omega_0}{2}\left(\tau\omega_0 \pm \sqrt{\tau^2\omega_0^2 - 4}\right)$. Similarly, for the pole we get, $\omega_p = j\frac{\omega_0}{2}\left(\tau\omega_0 \mp \sqrt{\tau^2\omega_0^2 - 4}\right)$. Depending upon the value of $\tau\omega_0$, we obtain three scenarios: $\tau\omega_0 > 2$, when the pole-zero pair is symmetrically distributed along the imaginary frequency axis (over coupled regime), $\tau\omega_0 = 2$, when the pole-zero pair coincides with the $\omega_0$ but on the imaginary frequency axis with zero real part (critically coupled), and $\tau\omega_0 < 2$, when the pole-zero pair have nonzero real and imaginary parts (under-coupled). In the latter case, the negative (positive) sign in $\omega_z$ ($\omega_p$) represents the zero (pole) in the negative half-space which will be omitted, as it is a mirror reflection of the zero (pole) with a positive-valued real part of the frequency. The case $\tau\omega_0 < 2$ is of utmost importance for applications, as here, the zero and pole possess nonzero real frequencies, making them accessible to waves and signals. We will assume this scenario in the following analysis.

Now, assuming a port weakly coupled to the resonator ($\tau\omega_0 \to 0$) and calculating the reflection coefficient in the complex frequency plane reveals a pole ($|r| \to \infty$), and a corresponding complex-conjugated zero ($|r| \to 0$) located very close to the same point on the real frequency axis ($\omega_0$). In the extreme scenario where the coupling is absent, $\tau\omega_0 = 0$, the convergence of the pole and zero on the real axis occurs, leading to the formation of a discrete bound state characterized by a diverging quality factor (Q-factor). Increasing the circuit's coupling to the port (increase of $\tau\omega_0$) causes the pole and zero to move further apart in the complex plane, **Figure 2(b)**. A deeply located zero in the complex plane implies that a circuit excited by a real frequency (CW) signal or a spectrum of signals will mostly reflect, indicating a mismatch with the port.

## 2. Non-oscillatory-to-oscillatory signal conversion in a detuned system

In this section, we provide additional insights into the conversion efficiency of the circuits with varying pole configurations. As discussed in the main text, the efficiency is highly dependent on the coupling between the poles in the complex plane. Figure S1 shows the configuration and results for a circuit with detuned poles, where the poles have different real frequencies. In this setup, the coupling is less optimal compared to the matched case. The efficiency analysis reveals that the conversion efficiency drops to approximately 20%. This decrease highlights the importance of precise tuning in achieving higher efficiency. For comparison, Figure 3 illustrates the efficiency of a circuit with matched poles, where the poles have similar real frequencies. As discussed in the main text, this configuration achieves a higher conversion efficiency of 31%. The improved efficiency is due to the optimal coupling of the poles, which allows better energy transfer and reduced losses.

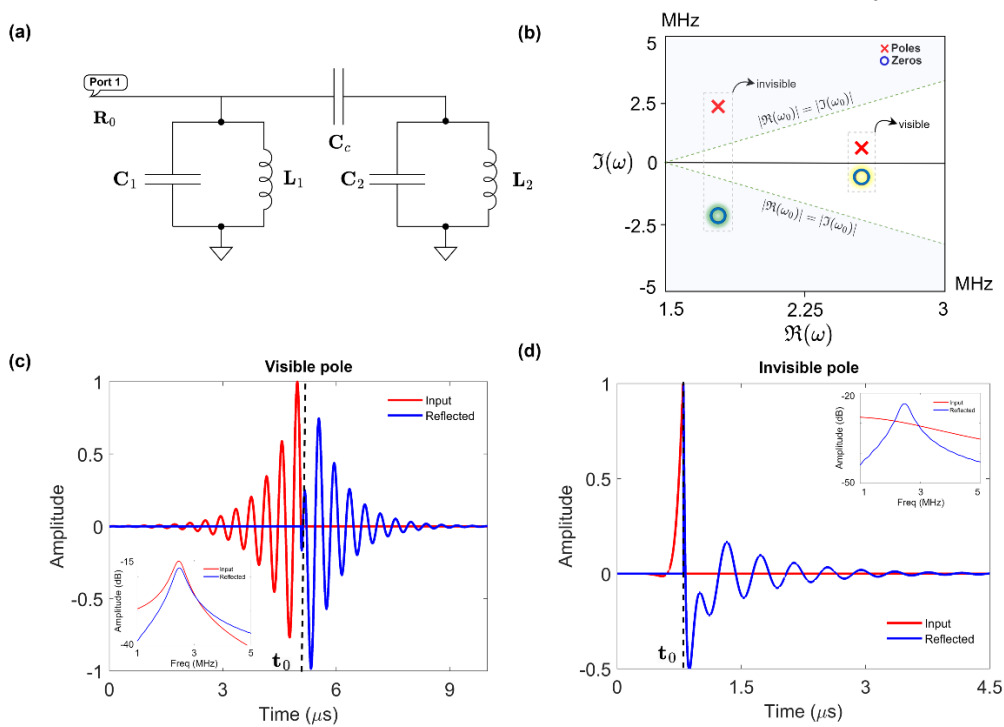

*Figure S5.* (a) Circuit with two resonators: the left (invisible pole) resonator is coupled to the port, and the right has a visible pole. (b) The real frequencies of the poles are mismatched. (c) Excitation via the visible pole zero shows no scattering before cut-off and a reflected signal after cut-off. The output signal is almost time-reversed. The inset shows the FFT spectrum centered around 2.5 MHz for both signals. (d) Excitation at the invisible pole's zero complex frequency results in an oscillating reflected signal after cut-off. The FFT spectrum reveals the visible pole's 2.5 MHz frequency, demonstrating energy coupling from the invisible to the visible pole. The efficiency of the non-oscillatory-to-oscillatory signal conversion is ~20%.

## 3. Non-oscillatory-to-oscillatory signal conversion with imaginary frequency signals

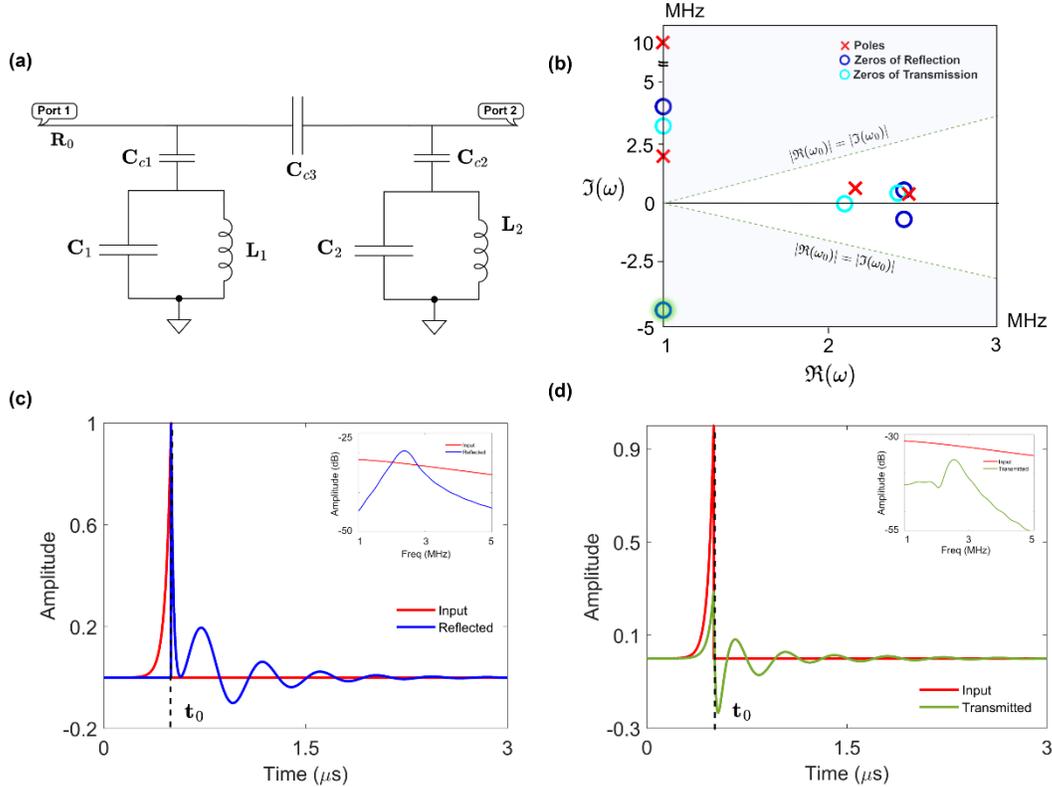

*Figure S6*. (a) Schematic of the 2-port circuit with parameters: $L_1 = L_2 = 4$ µH, $C_1 = C_2 = 0.4$ nF, $C_{c1} = C_{c2} = 1$ nF, and $Cc3 = 0.5$ nF. (b) Reflection coefficient in the complex frequency plane showing poles on both the complex plane and the imaginary axis. The zero of excitation is highlighted in green. (c) Reflected signal upon excitation by a purely imaginary pulse. (d) Transmitted signal under the same conditions. The computed conversion efficiency of non-oscillating to oscillating radiation is approximately 38%.

Here, we present the analysis of a 2-port circuit with the following parameters: $L_1 = L_2 = 4$ µH, $C_1 = C_2 = 0.4$ nF, $C_{c1} = C_{c2} = 1$ nF, and $C_{c3} = 0.5$ nF. Figure S2(a) illustrates the circuit diagram. Our analytical investigation of the reflection coefficient in the complex frequency plane reveals the presence of poles not only in the complex plane but also on the imaginary axis, as depicted in Figure S2(b). This behavior is characteristic of the 2-port regime, where the poles and zeros of the reflection are not complex conjugates due to radiation to the second port. In contrast, the 2x2 Scattering matrix for this circuit exhibits complex conjugate poles and zeros.

Figures S2(c) and S2(d) show the reflected and transmitted signals, respectively, when the circuit is excited by a *purely imaginary pulse*. The zero of excitation is highlighted in green in Figure S2(b). The conversion efficiency from non-oscillating to oscillating radiation is approximately 38%, slightly higher than in the single-port regime due to the presence of two radiation channels, although the poles are largely detuned.

## 4. Coupled Mode Theory for a 2 Resonator System

In this section, we derive the steady-state solution for a system consisting of two coupled resonators. The first resonator is coupled to a waveguide, and the second resonator is coupled to the first resonator. We consider the effect of an external excitation pulse with a complex frequency, allowing for the possibility of exponential growth or decay. The system is described by the following coupled differential equations(3–5). For the first resonator:

$$\frac{da_1(t)}{dt} = \left(i\omega_{01} - \frac{1}{\tau_{o1}} - \frac{1}{\tau_{e1}}\right)a_1(t) + \sqrt{\frac{2}{\tau_e}}S_{ex}(t) + ga_2(t) \tag{1}$$

For the second resonator:

$$\frac{da_2(t)}{dt} = \left(i\omega_{02} - \frac{1}{\tau_{o2}}\right)a_2(t) + ga_1(t), \tag{2}$$

where $a_1(t)$ and $a_2(t)$ are the mode amplitudes of the first and second resonators, respectively, $\omega_{01}$ and $\omega_{02}$ are the eigenfrequencies of the first and second resonators, respectively, $\tau_{o1}$ and $\tau_{o2}$ are the intrinsic decay times of the first and second resonators, respectively, $\tau_{e1}$ is the coupling time of the first resonator to the waveguide, $g$ is the coupling coefficient between the two resonators, $S_{ex}(t)$ is the external excitation pulse.

**Steady-State Solution**

We assume the external excitation pulse $S_{ex}(t)$ is a wave with a complex frequency $(\omega_r + i\omega_i)$, which can be expressed as: $S_{ex}(t) = S_0 e^{(i\omega_r - \omega_i)t}$. We seek steady-state solutions of the form: $a_1(t) = A_1 e^{(i\omega_r - \omega_i)t}$, $a_2(t) = A_2 e^{(i\omega_r - \omega_i)t}$. Substituting these forms into the governing equations and simplifying, we obtain the following matrix equation:

$$\begin{pmatrix} i\omega_r - \omega_i - i\omega_{01} + \frac{1}{\tau_{o1}} + \frac{1}{\tau_{e1}} & -g \\ -g & i\omega_r - \omega_i - i\omega_{02} + \frac{1}{\tau_{o2}} \end{pmatrix} \begin{pmatrix} A_1 \\ A_2 \end{pmatrix} = \begin{pmatrix} \sqrt{\frac{2}{\tau_{e1}}}S_0 \\ 0 \end{pmatrix}$$

Let: $M = \begin{pmatrix} i\omega_r - \omega_i - i\omega_{01} + \frac{1}{\tau_{o1}} + \frac{1}{\tau_{e1}} & -g \\ -g & i\omega_r - \omega_i - i\omega_{02} + \frac{1}{\tau_{o2}} \end{pmatrix}$. The solution is: $\begin{pmatrix} A_1 \\ A_2 \end{pmatrix} = M^{-1} \begin{pmatrix} \sqrt{\frac{2}{\tau_{e1}}}S_0 \\ 0 \end{pmatrix}$.

The determinant of $M$ is: $\det(M) = \left(i\omega_r - \omega_i - i\omega_{01} + \frac{1}{\tau_{o1}} + \frac{1}{\tau_{e1}}\right)\left(i\omega_r - \omega_i - i\omega_{02} + \frac{1}{\tau_{o2}}\right) - g^2$. Thus, the steady-state solutions are:

$$a_1(t) = A_1 e^{(i\omega_r - \omega_i)t} = \frac{\left(i\omega_r - \omega_i - i\omega_{02} + \dfrac{1}{\tau_{o2}}\right)\sqrt{\dfrac{2}{\tau_e}} S_0}{\det(M)} e^{(i\omega_r - \omega_i)t}$$

$$a_2(t) = A_2 e^{(i\omega_r - \omega_i)t} = \frac{g\sqrt{\dfrac{2}{\tau_e}} S_0}{\det(M)} e^{(i\omega_r - \omega_i)t}$$

The reflection coefficient $r$ from the first resonator is defined as the ratio of the reflected wave to the incident wave at the waveguide: $S_{\text{ref}}(t) = -S(t) + \sqrt{\dfrac{2}{\tau_e}} a_1(t)$:

$$r(\omega_r, \omega_i) = \frac{S_{\text{ref}}(t)}{S(t)} = 1 - \frac{2}{\tau_e} \frac{\left(i\omega_r - \omega_i - i\omega_{02} + \dfrac{1}{\tau_{o2}}\right)}{\det(M)} \tag{3}$$

This completes the derivation of the coupled mode theory for the two resonator system and the steady-state solution. We used Eq.(3) to plot the results in Figure 4(b).

The results in Figures 4(c) and (d) were obtained by solving the Eq.(1)-(2) numerically.

## REFERENCES

1. D. M. Pozar, *Microwave Engineering, 4th Edition* (John Wiley & Sons, Inc., 2011).

2. A. V. Marini, D. Ramaccia, A. Toscano, F. Bilotti, Perfect Matching of Reactive Loads Through Complex Frequencies: From Circuital Analysis to Experiments. *IEEE Trans. Antennas Propag.* **70**, 9641–9651 (2022).

3. W. Suh, Z. Wang, S. Fan, Temporal coupled-mode theory and the presence of non-orthogonal modes in lossless multimode cavities. *IEEE J. Quantum Electron.* **40**, 1511–1518 (2004).

4. H. A. Haus, W. Huang, Coupled-mode theory. *Proc. IEEE* **79**, 1505–1518 (1991).

5. H. A. Haus, *Waves and Fields in Optoelectronics* (Prentice-H, Englewood Cliffs, NJ, Englewood Cliffs, NJ, Prentice-H., 1984).